\documentclass[aps,prl,twocolumn, showpacs, superscriptaddress, longbibliography]{revtex4-1}
\usepackage{graphicx}

\begin{document}

\title{Quantum manipulation of biphoton spectral distributions in a 2D frequency space toward arbitrary shaping of a biphoton wave packet}
\author{Rui-Bo Jin}
\affiliation{Laboratory of Optical Information Technology, Wuhan Institute of Technology, Wuhan 430205, China\\}
\author{Ryoji Shiina}
\affiliation{The University of Electro-Communications, 1-5-1 Chofugaoka, Chofu, Tokyo, Japan}
\author{Ryosuke Shimizu}
\email{r-simizu@uec.ac.jp}
\affiliation{The University of Electro-Communications, 1-5-1 Chofugaoka, Chofu, Tokyo, Japan}

\date{\today }

\begin{abstract}
In this work, we experimentally manipulate the spectrum and phase of a biphoton wave packet in a two-dimensional frequency space.  The spectrum is shaped by adjusting the temperature of the crystal, and the phase is controlled by tilting the dispersive glass plate. The manipulating effects are confirmed by measuring the two-photon spectral intensity (TSI) and the Hong-Ou-Mandel (HOM) interference patterns. Unlike the previous  \emph{independent manipulation}  schemes, here we perform  \emph{joint manipulation}  on the biphoton spectrum.
The technique in this work paves the way for arbitrary shaping of a multi-photon wave packet in a quantum manner.
\end{abstract}



\maketitle

\section{Introduction}
The ability to generate, manipulate and measure entangled photon pairs (biphotons) with an arbitrary time-frequency quantum state is of vital importance not only for developing a tool for practical quantum information and communication technologies, but also for conducting fundamental photonics studies \cite{Shih2011}.
Recently, the time-frequency duality of biphotons has been demonstrated by measuring two-photon spectral intensity (TSI) and two-photon temporal intensity (TTI) in the same condition \cite{JinShimizu2018arXiv}, an important step for generating and measuring biphoton wave packets in a time-frequency domain.
The manipulation of biphoton wave packets must be mastered before the arbitrary shaping of a biphoton wave packet in two-dimensional (2D) time-frequency space can be achieved.
Especially, it is necessary to create multiple frequency modes in a 2D frequency space and to control the relative phases between the frequency modes \cite{Chan2011}.

Many previous works have been dedicated to the manipulation of the spectrum and phase of a biphoton wave packet.
For narrow-band biphotons generated from a four-wave-mixing (FWM) process in a cold atomic ensemble within a
magneto-optical trap (MOT), the biphoton temporal-spectral distribution is shaped by modulating the
pump lasers using electro-optical modulators (EOM) \cite{Chen2010}, by using a spatial light modulator (SLM) \cite{Zhao2015}, or by manipulating
the dispersive properties of the electromagnetically induced transparency medium \cite{Cho2014}.
For biphotons generated from a spontaneous parametric downconversion (SPDC) process in a nonlinear crystal, the shaping of the wavepacket has been realized by several methods \cite{Torres-Company2009, Tischler2015, Su2016, Ansari2018, Valencia2007, Mittal2017, Matsuda2016, Peer2005, Bernhard2013}.
For example, Valencia et al. controlled the joint spectra of biphotons by spatially shaping the pump beam using a hologram \cite{Valencia2007};
Mittal et al. magnified biphoton's temporal distribution using a time lens composed of dispersion fibers, EOM and gratings \cite{Mittal2017};
Matsuda shaped  biphotons using cross-phase modulations with a photonic crystal fiber (PCF) \cite{Matsuda2016};
and the  spectra and phases of biphotons can also be manipulated using SLM \cite{Peer2005, Bernhard2013}.

Although the previous schemes \cite{Chen2010, Zhao2015, Cho2014, Torres-Company2009, Tischler2015, Su2016, Ansari2018, Valencia2007, Mittal2017, Matsuda2016, Peer2005, Bernhard2013} have provided several effective ways to modulate biphotons,
these techniques rely on preexisting technologies to \emph{independently modulate} the constituent photons  of a biphoton wave packet.
In this work, we  propose and demonstrate another approach that \emph{jointly modulates} a biphoton wave packet other than two photons independently.
The spectra of the biphotons are shaped by simply adjusting the temperature of the crystal and the phase is adjusted just by tilting the angle of a glass plate. The shaping effect is confirmed by measuring the TSI and Hong-Ou-Mandel (HOM) \cite{Hong1987} interference patterns.

\section{Experiment}
%
\begin{figure}[!bhp]
\centering
\includegraphics[width= 0.45\textwidth]{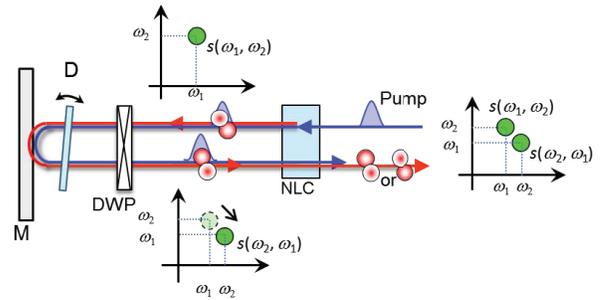}
\caption{Schematic of creation of  a discrete-frequency entangled biphoton in bidirectional pump. NLC, nonlinear crystal; DWP, dual-wavelength wave plate; M, mirror; D, dispersive medium.}
\label{scheme}
\end{figure}
%

%
%
\begin{figure*}[tbhp]
\centering
\includegraphics[width= 0.95\textwidth]{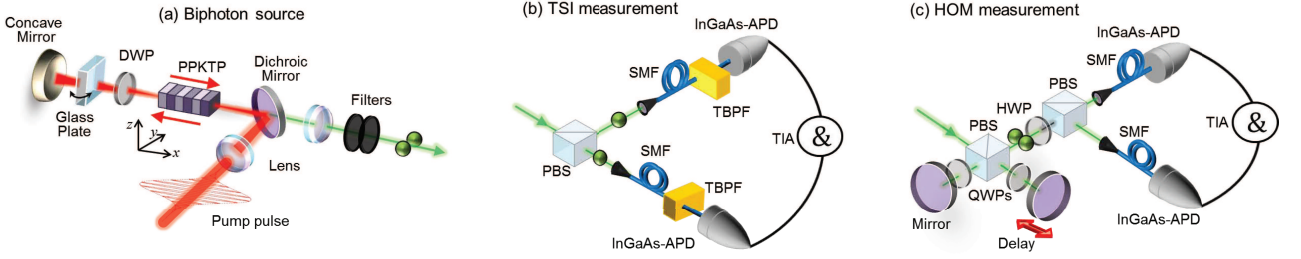}
\caption{Experimental setup for (a) biphoton source, (b) TSI measurement, (c) HOM measurement. DWP, dual-wavelength wave plate; SMF, single-mode fiber; APD, avalanche photodiode; PBS, polarizing beam splitter; HWP, half-wave plate.}
\label{setup}
\end{figure*}

The experimental scheme for the generation of an entangled photon source is shown in Fig.\,\ref{scheme}.
The pump laser pulses pass through a nonlinear crystal (NLC), a dual-wavelength wave plate (DWP), and a dispersive medium (D) twice so as to generate an entangled state with spectrum of
\begin{equation}\label{eq1}
S(\omega_1, \omega_2)=s(\omega_1, \omega_2)+e^{i \phi}s(\omega_2, \omega_1).
\end{equation}
Here, the $s(\omega_1, \omega_2)$ state is generated when the pump passes though the type-II  NLC; $\omega_1$ and $\omega_2$ are the angular frequencies of the down converted biphotons; the relative phase $\phi$ is introduced by the dispersive medium, which has different chromatic dispersions for the pump and the downconverted daughter photons. $\phi$ can be continuously controlled by tilting the dispersive medium.

Note that the DWP works as a half-wave plate for the pump and a quarter-wave plate for the biphoton. As a result of the passing of the DWP twice, the polarization of the pump pulse is not changed while the polarizations of the constituent photons of the biphoton are interchanged, i.e., $s(\omega_1, \omega_2)$ is changed to $s(\omega_2, \omega_1)$. Further, the relative phase $\phi$  can maintain long-term stability against the disturbance, since the pump pulses propagate together with the biphotons between the NLC and the mirror M.

The real experimental setup is shown in Fig.\,\ref{setup} (a-c). Figure\,\ref{setup}(a) is the setup for the entangled photon source based on the scheme in Fig.\,\ref{scheme}. Here, the group-velocity-matched (GVM) periodically poled KTiOPO$_4$ (PPKTP) crystal  is adopted as the NCL for type-II SPDC (y$\to$ y + z), and  $\omega_{1(2)}$   corresponds to the polarization aligned along the y- (z-) axis \cite{Konig2004, Shimizu2009}.
A glass plate is employed as the dispersion medium.
The pump pulses from a mode-locked Ti:sapphire laser have a center wavelength of 792 nm  and a bandwidth of approximately 0.2 nm with a repetition  rate of 76 MHz.
Figure\,\ref{setup}(b) is the setup for the two-photon spectral intensity (TSI) measurement, similar to what we used in Refs. \cite{Shimizu2009, Jin2013OE}.
Figure\,\ref{setup}(c) shows the HOM interference measurement; more technical details can be found in Refs. \cite{Bisht2015, Jin2018optica}.

\section{Results}

%
\begin{figure}[tbhp]
\centering
\includegraphics[width= 0.49\textwidth]{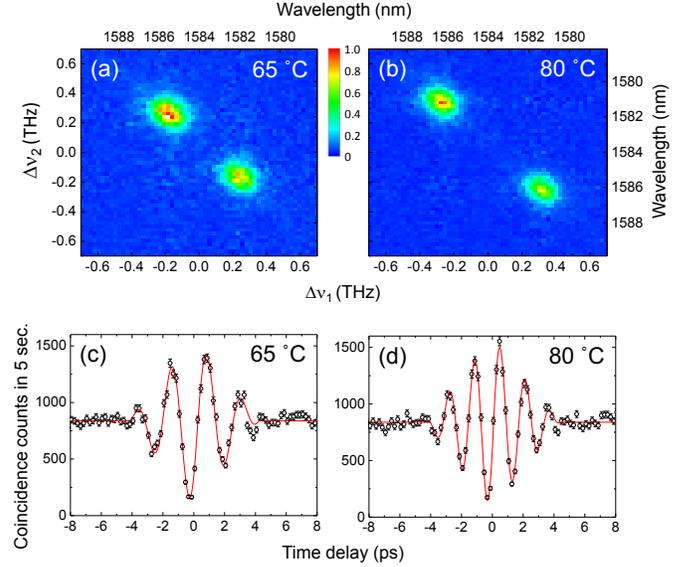}
\caption{Observed two-photon spectral intensities with normalization (upper) and their HOM interference patterns (lower). The plots on the left (right) show  the data at the PPKTP temperature of 65\,$^{\circ}$C (80\,$^{\circ}$C).   Error bars are equal to the square root of each data point by assuming Poissonian counting statistics.}
\label{spectrum}
\end{figure}

First, we demonstrate the shaping of the biphoton spectrum by adjusting the temperature of the PPKTP crystal. Temperature is an important parameter for manipulating the spectra of biphotons from periodically poled crystals \cite{Fedrizzi2009, Zhou2018arXiv}. The effect of the shaping is confirmed by measuring the TSI and HOM interference.
Figure \,\ref{spectrum} (a) shows the measured TSI at  65\,$^{\circ}$C for PPKTP.
Note that $\Delta \nu _{1(2)}$  in Fig.\,\ref{spectrum} is the shifted frequency for $\omega _{1(2)}$ from the center frequency.
Thanks to the GVM condition, the states $s(\omega_1, \omega_2)$ and $s(\omega_2, \omega_1)$ in Eq.(1) have  round shapes in the TSI  \cite{Yabuno2012, Bruno2014,  Jin2016PRAppl, Jin2017PRA, Greganti2018} .
The  two frequency modes were separated  by 0.40 THz in Fig.\,\ref{spectrum} (a).
When the temperature was increased to 80\,$^{\circ}$C in Fig.\,\ref{spectrum}(b), the frequency separation was 0.61 THz.
Figure \,\ref{spectrum} (c) shows the HOM interference patterns at 65\,$^{\circ}$C with an interference visibility of 84.6$\pm 2.0\%$ and a beating period of 2.3 ps. Here, the  visibility is defined as (max-min)/(max+min).
Higher visibilities in the HOM patterns guarantee the exchange symmetry of the two-photon spectral distribution.
At 80\,$^{\circ}$C in Fig.\,\ref{spectrum}(d), the interference visibility was 88.2$\pm 1.7\%$,  comparable to that in Fig.\,\ref{spectrum}(c), but the beating period  decreased to 1.6 ps. The temporal period was shorter in Fig.\,\ref{spectrum}(d)   because the spectral period was enlarged in Fig.\,\ref{spectrum}(b). The frequency separations in the TSI data are in good agreement with the beat frequencies in the HOM patterns, since $\Delta \upsilon = 1 / \Delta \tau$, where $\Delta \upsilon$ is the separation of the spectral modes and $\Delta \tau$ is the beating period.

\begin{figure}[tbhp]
\centering
\includegraphics[width= 0.45\textwidth]{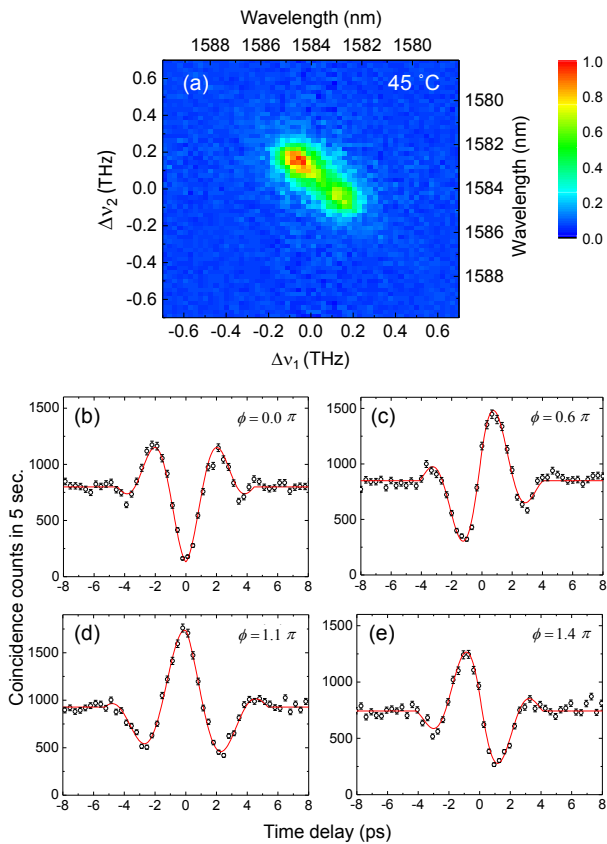}
\caption{Demonstration of the controllability of the relative phase between the two spectral modes $s(\omega_1, \omega_2)$ and $s(\omega_2, \omega_1)$. (a) Normalized two-photon spectral intensity (TSI) at  45\,$^{\circ}$C for PPKTP . (b)-(e) Corresponding HOM interference patterns at different relative phases. Error bars  assume Poissonian counting statistics.
}
\label{phase}
\end{figure}

Next, we demonstrate controlling the relative phase $\phi$ by tilting the glass plate.
We fixed the temperature at 45\,$^{\circ}$C for the PPKTP with the measured TSI shown in Fig.\,\ref{phase}(a).
With a phase at $\phi=0.0$ $\pi$, the HOM interference pattern was symmetric as shown in Fig.\,\ref{phase}(b), with a visibility of 83.6$\pm 2.9\%$ and a beating period of 4.4 ps.
When the phase was increased to 0.6 $\pi$, the HOM interference pattern was asymmetric, as shown in Fig.\,\ref{phase}(c). The phase can be continuously increased, and Figs.\,\ref{phase}(c,d) show the case of $\phi$ = 1.1 $\pi$ and 1.4 $\pi$.
All the patterns in Fig.\,\ref{phase}(b-e) have the same period.

%

\section{Discussion}
As shown in Fig.\,\ref{spectrum}, we have created discrete frequency modes in a 2D frequency space. The separation between the two frequency modes can be controlled just by changing the crystal's temperature.
To the best of our knowledge, this is the first experimental demonstration of the relationship between the discrete frequency mode in the TSI and the HOM interference patterns.

With the data shown in Fig.\,\ref{phase}, we have controlled the relative phase between two frequency modes by adjusting the tilting angle of the glass plate.
This technique will be useful for future applications requiring precise control of the relative phase in a time-frequency entangled state.
For the maximal controlling range of our technique, the spectral separation can reach 2.4 THz by heating the crystal from room temperature to 200\,$^{\circ}$C;  by tilting the glass plate,  the phase can be adjusted from 0 to 2 $\pi$.

The next step in our work is to verify the temporal shaping effect of a biphoton wave packet with the quantum manipulation technique, which has been presented in this study by measuring a two-photon temporal distribution.
Our approach to time-frequency entangled photons may open up new directions for optical science and technologies. In particular, the technique for shaping a biphoton wave packet could be developed as quantum optical synthesis in a 2D time-frequency space,  which might be a next-generation technology beyond the conventional optical synthesis  carried out in 1D systems.

\section{Conclusion}
We have demonstrated  \emph{joint manipulation} of the biphoton spectrum and phase. The frequency separation can achieve 2.4 THz by changing the crystal temperature, and the phase can be adjusted by 2 $\pi$ by tilting the angle of a glass plate. We have also experimentally demonstrated, for the first time, the correspondence between TSI and HOM interference with different spectra and phases. Our work opens up new possibilities for shaping entangled photons for various quantum optical and quantum information applications.

\section*{Acknowledgments}
R.J. is supported by the National Natural Science Foundations of China (Grant No.11704290), and by a fund from the Educational Department of Hubei Province, China (Grant No. D20161504). R.S. acknowledges support from the JSPS KAKENHI Grant Number JP17H01281, the Research Foundation for Opto-Science and Technology, Hamamatsu, Japan, and support from the Matsuo Foundation, Tokyo, Japan.

\end{document}